\documentclass[twocolumn,showpacs,preprintnumbers,amsmath,amssymb]{revtex4}
\usepackage{graphicx}
\usepackage{dcolumn}% Align table columns on decimal point
\usepackage{bm}% bold math

\begin{document}
%%% ----------------------------------------------------------------------
\title{Stable Spatial Langmuir Solitons.}

\author{T.A. Davydova}
 \email{tdavyd@kinr.kiev.ua}
\author{A.I. Yakimenko\footnote{also at Department of Physics, Kiev University, prosp.
Glushkova 6, Kiev 03022, Ukraine}}%
\email{ayakim@kinr.kiev.ua}
\author{Yu. A. Zaliznyak}
\email{zalik@kinr.kiev.ua} \affiliation{Institute for Nuclear
Research, Prospect Nauki 47, Kiev 03680, Ukraine}
\date{\today}
\begin{abstract}
We study localized two- and three-dimensional Langmuir solitons in
the framework of model based on generalized nonlinear
Schr\"odinger equation that accounts for local and nonlocal
contributions to electron-electron nonlinearity. General
properties of solitons are investigated analytically and
numerically. Evolution of three-dimensional localized wave packets
has been simulated numerically. The additional nonlinearities are
shown to be able to stabilize both azimuthally symmetric
two-dimensional and spherically symmetric three-dimensional
Langmuir solitons.

\end{abstract}
\maketitle
Starting from the early 70-th, the fundamental role of Langmuir
solitons (LS) in strong plasma turbulence is commonly known
\cite{Zakharov1972, Rudakov1974}. Formation of LS are connected
with the plasma extrusion from the regions of strong
high-frequency electric field and trapping of plasmons into the
formed density well (caviton). However, in the previous simplified
theoretical models, two-dimensional (2D) and 3D solitons occurs to
be unstable with respect to a collapse: above some threshold
power, the size of the structure shrinks infinitely, forming
density singularity at finite time \cite{Zakharov1972}.
Physically, when the size of collapson approaches few Debye radii
$r_D$, it should damp rapidly which may result in fast plasma
heating. Nevertheless, experimental observations
\cite{AntipovNezlin1981, Tran82, Eggleston1982, Wong1984,
Wong1985} of quasi-2D and 3D Langmuir collapse demonstrate
saturation of wave-packet's spatial scale at some minimum value
being of order $(10 \div 40)r_D$. It have been observed in Refs.
\cite{Wong1984, Wong1985} that at times $t>50\omega_{Pi}^{-1}$,
where $\omega_{Pi}$ is the ion plasma frequency, Langmuir wave
packets show considerably slow dynamics (subsonic regime). To our
best knowledge, these observations do not meet an appropriate
theoretical explanation yet.

Various additional linear and nonlinear effects, such as
higher-order dispersion \cite{Karpman96, Zakharov98,OurPRE03}, the
saturation of nonlinearity \cite{ZakharovSobolevSynakh1971,
VakhitovKolokolov,OurPRE03}, nonlocal wave interaction
\cite{DavydovaFischuk95, BangKrolikowskiWyllerRasmussen2002,
Parola1998, KrolikovskiBangRasmussen2001, Brizhik2001,
ZakrzewskiUFZ03, PerezGarciaKonotopPRE2000}, may arrest wave
collapse both in 2D and in 3D
\cite{Karpman96,VakhitovKolokolov,Zakharov98,DavydovaFischuk95,
BangKrolikowskiWyllerRasmussen2002}. As it is shown in
\cite{Kuznetsov1976}, the local part of electron-electron
nonlinearity (resulting from the interaction with the second
harmonic) counteracts the contraction of wave packet. At the same
time, the nonlocal contribution of the additional nonlinear term
was omitted in \cite{Kuznetsov1976}, though it is of great
importance for sufficiently narrow and intense wave packets. In
this Letter we take into consideration both these extra nonlinear
effects. As it will be shown below, the role of nonlocal
nonlinearity is quantitatively even more significant.  The
nonlocal nonlinearity is of great importance not only when
describing soliton formation in plasmas \cite{DavydovaFischuk95,
PorkolabGoldman76, OurUFZh03}, but also in the theory of
Bose-Einstein condensates or matter waves \cite{Parola1998,
PerezGarciaKonotopPRE2000, KrolikovskiBangRasmussen2001,
BangKrolikowskiWyllerRasmussen2002}, and in the construction of an
adequate continuum model of the electron-phonon interaction in
discrete 2D and 3D lattices \cite{Brizhik2001, ZakrzewskiUFZ03}.

The evolution of radial component of electric field strength $E$
of a Langmuir 2D and 3D wave packet is described by the set of
equations:
%%%
\begin{eqnarray}
\nonumber \textrm{i}\frac{\partial E}{\partial t} +
  \frac{3}{2}\omega_p r_D^2 \frac{\partial}{\partial r}
  r^{1-d}\frac{\partial }{\partial r} r^{d-1}E - \frac{\omega_p}{2}\frac{n_1}{n_0} E
\\
 \nonumber -
  \frac{1}{48\pi m n_0 \omega_p}\frac{E \left| E
  \right|^2}{r^2}
  + \frac 3 2 \,\, \frac{E \Delta_r\left|E\right|^2}{48\pi m n_0 \omega_p}=0,\\
\nonumber \left( \frac{\partial^2}{\partial t^2} - c_S^2 \Delta_r
  \right) \left( n_1 + \frac{\Delta_r \left|E \right|^2}{16 \pi m \omega_p^2 }
  \right) = \frac{\Delta_r \left| E \right|^2}{16\pi M},
\end{eqnarray}
where azimuthal or spherical symmetry is supposed, $r$ -- radial
coordinate, $\Delta_r=\frac{\partial^2}{\partial
r^2}+\frac{d-1}{r}\frac{\partial}{\partial r}$,  $r_D$ -- Debye
radius, $\omega_p$ -- electron plasma frequency, $d$ is the number
of space dimensions, $m$ and $M$ are the electron and ion masses,
$n_0$, $T_e$ -- background electron density and temperature
respectively and $c_S=\sqrt{T_e/M}$ is the ion sound speed, $n_1$
is the electron density perturbation. This equation set is valid
if $W/nT<(kr_D)^{-1}$, where $W=E^2/8\pi$, $k$ being the effective
wave number of the packet. In \cite{Kuznetsov1976}, the similar
set of equations for Langmuir wave field was derived, however, the
nonlocal nonlinear terms were omitted.

%---------------------------
\begin{figure*}[hbt]
\begin{centering}
\includegraphics[width=\textwidth]{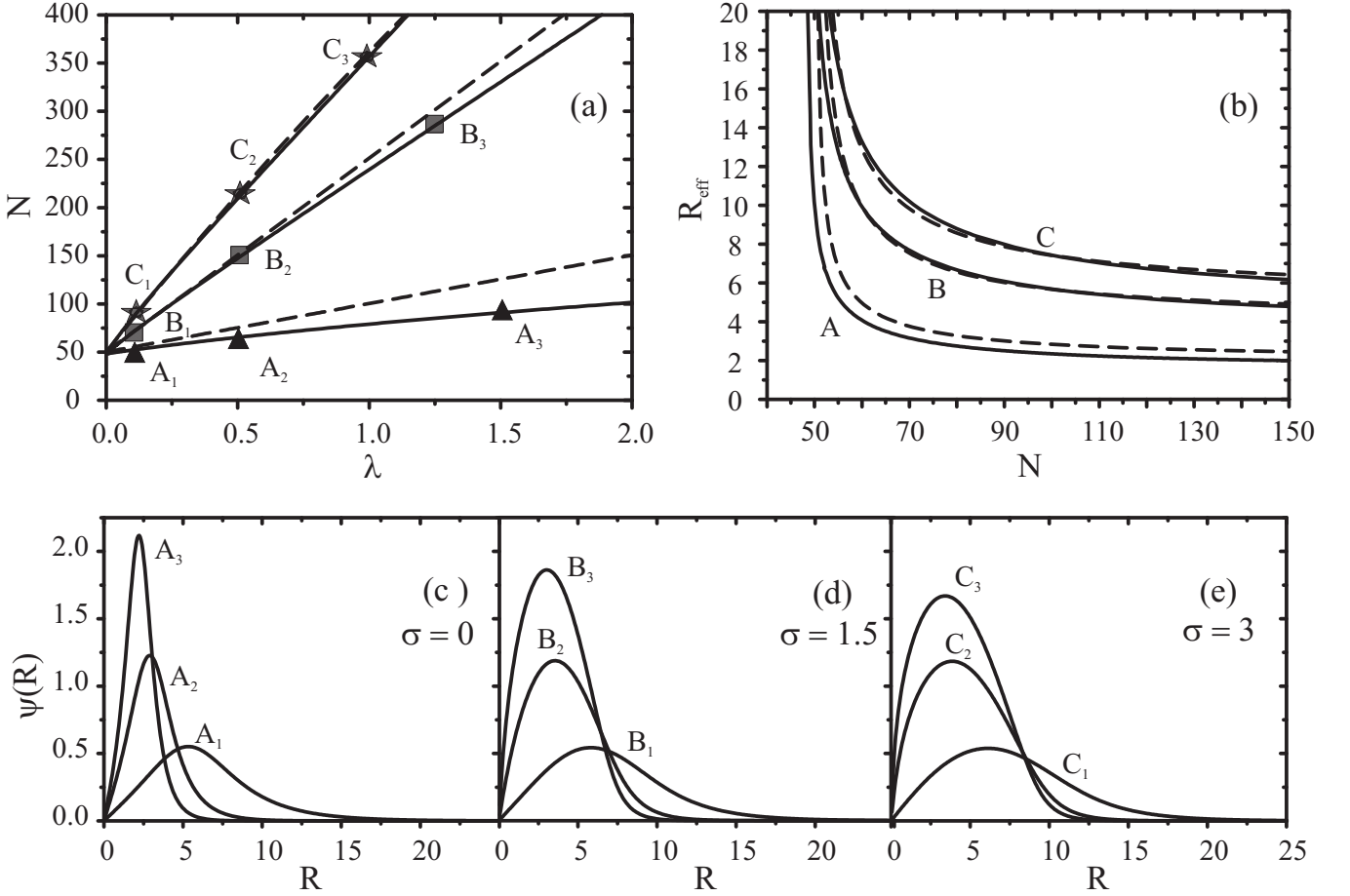}
 \caption{Two-dimensional Langmuir solitons: (a) EDD for
 $\sigma=0$ (A), $\sigma=1.5$ (B), and $\sigma=3.0$ (C);
 (b) Effective soliton width
 vs plasmon number;
 (c)-(e) soliton profiles for different $\sigma$. Each profile
 corresponds to the point marked at the EDD lines.}
  \label{Fig2d}
\end{centering}
\end{figure*}
%---------------------------
\begin{figure*}[hbt]
\begin{centering}
\includegraphics[width=\textwidth]{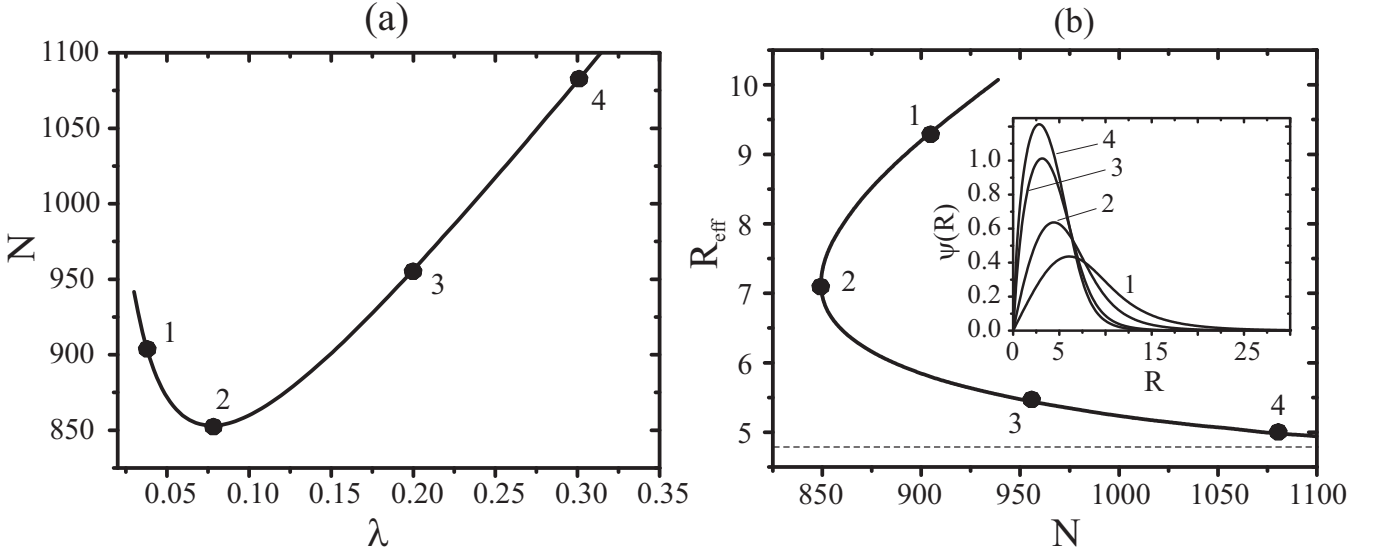}
 \caption{Three-dimensional Langmuir solitons. (a) Plasmon number $N$ vs
 nonlinear frequency shift $\lambda$ for
 $\sigma=1.5$;
  (b) Effective soliton
 width
 vs plasmon number. The inset: soliton profiles
 corresponding to the points marked at the
 $N(\lambda)$ and $R_{\textrm{eff}}(N)$ curves. }
  \label{Fig3d}
\end{centering}
\end{figure*}
%%%

We consider subsonic motions and neglect the terms with time
derivatives in the second equation of above set.  As result, this
set is reduced to the single partial differential equation:
%%%
\begin{eqnarray}
\nonumber \textrm{i}\frac{\partial E}{\partial t} +
D\frac{\partial}{\partial r}\,
  r^{1-d}\frac{\partial }{\partial r} r^{d-1}E + B E \left| E
\right|^2 \\
+ C E \Delta_r \left| E \right|^2 - \Gamma \frac{E
\left| E \right|^2}{r^2} = 0, \label{GNSE2D3D}
\end{eqnarray}
where  the coefficients $D$, $B$, $C$, $\Gamma$ are given by the
expressions:
\begin{eqnarray}
\nonumber
D=\frac 32  \omega_p r_D^2, \,\,\, B=\frac{\omega_p}{32 \pi M n_0
c_S^2 }, \\
\nonumber
C=\frac{7}{96 \pi m n_0 \omega_p}, \,\,\, \Gamma=\frac{1}{48 \pi m
n_0 \omega_p}.
\label{DBCGCoefs}
\end{eqnarray}
Equation (\ref{GNSE2D3D}) is closely related to nonlinear
Schr\"odinger equation with modified linear term (proportional to
$D$). The nonlinear part of this equation includes common cubic
nonlinearity (term proportional to $B$) as well as nonlocal
 (term proportional to $C$) and local
 (term with $\Gamma$) parts of electron-electron nonlinearity.

Equation (\ref{GNSE2D3D}) conserves the following integrals: the
plasmon number
\begin{equation}
 \label{PlNum2D}
 N = \int\left| E \right|^2 d \textbf{r},
\end{equation}
%%%%%
and Hamiltonian:
\begin{eqnarray}
 \nonumber
 H = D \int\left|r^{1-d}\frac{\partial }{\partial r} r^{d-1}E\right|^2d\textbf{r}
  -\frac B2\int|E|^4d\textbf{r}
\\
  +\frac C2 \int \left( \nabla \left| E \right|^2 \right)^2
d\textbf{r}
  + \frac \Gamma 2 \int \frac{\left| E \right|^4}{r^2}
  d\textbf{r}.
  \label{Hamilt2D}
\end{eqnarray}
%%%
Other integrals (momentum and angular momentum) are equal to zero
in the case under consideration.

Let us show that the effective width $r_{\textrm{eff}}$
\begin{equation}\label{eq:Reff}
r_{\textrm{eff}}^2=\frac{1}{N}\int r^2|E|^2 d\textbf{r},
\end{equation}
of any wave packet governed by the Eq. (\ref{GNSE2D3D}) is bounded
from below in the most interesting case of self-trapped wave
packets having negative Hamiltonian. If both $C$ and  $\Gamma$ in
Eq. (\ref{GNSE2D3D}) are equal to zero, and $H<0$, collapse
occurs. However, any of additional nonlinear terms (proportional
to $C$ or $\Gamma$) prevents collapse in 2D as well as in 3D
cases, or, in the other words, $r_\textrm{eff}$ is bounded from
below.

Let us define
\begin{displaymath}
r_1^2=\frac{\int r^2 \left| E \right|^4 d\textbf{r}}{\int \left| E
\right|^4 d\textbf{r}}, \,\,\, r_2^2=\frac{\int \left| E \right|^4
d\textbf{r}}{\int \left| E \right|^4 r^{-2}  d\textbf{r}},
\end{displaymath}
where all integrals are supposed to be finite. It is easy to show
that
%%%%%
\begin{equation}
r_1^2>r_2^2.
 \label{r1kv_GT_r2kv}
\end{equation}
%%%%%
Since Hamiltonian (\ref{Hamilt2D}) is assumed to be negative, it
is necessary that
\begin{equation}
 B \int \left| E \right|^4 d\textbf{r} >
 C \int \left( \nabla \left| E \right|^2 \right)^2 d\textbf{r}
 +  \Gamma \int r^{-2}\left|
 E \right|^4  d\textbf{r}.
 \label{B_Grater_C}
\end{equation}
Then we use the ``uncertanity relation" of the form
\begin{equation}
 r_1^2 \int \left( \nabla \left| E \right|^2 \right)^2 d\textbf{r} >
 \alpha(d)  \int \left| E \right|^4 d\textbf{r},
\label{Uncertanity}
\end{equation}
where $\alpha=3/4$ in 3D case and $\alpha=1/2$ in 2D case. One can
see from inequality (\ref{B_Grater_C}) that $r_2^2>\Gamma/B$,
thus, using Eq. (\ref{r1kv_GT_r2kv}) one gets $r_1^2>\Gamma/B$.
Taking into account that for a localized wave packet
$r_\textrm{eff}^2 \ge r_1^2$, we finally obtain
%%%%%
\begin{equation}
r_\textrm{eff}^2>\textrm{max} \left\{\frac \Gamma B, \,\,
\frac{\alpha(d)}{2B}\left[ C + \sqrt{C^2+\frac{4\Gamma
C}{\alpha(d)}} \right] \right\}.
 \label{Reff_Estimat}
\end{equation}
%%%%%
Thus, if $C\ne 0$ (or if $\Gamma \ne 0$), the wave packet can not
be contracted to the size smaller than $\sqrt{C/B}$ (or
$\sqrt{\Gamma/B}$).

By introducing the dimensionless variables:
\begin{eqnarray}
\nonumber R=\sqrt{\frac{3}{2}}\frac{r}{r_D}, \nonumber
\tau=\frac{9}{4}\omega_p t, \nonumber \Psi=E \sqrt{72\pi n_0 T_e},
\end{eqnarray}
Eq. (\ref{GNSE2D3D}) is reduced to
\begin{eqnarray}\nonumber
\textrm{i}\frac{\partial \Psi}{\partial \tau} + \frac{\partial^2
\Psi}{\partial R^2} + \frac{d-1}{R} \frac{\partial \Psi}{\partial
R } - \frac{d-1}{R^2} \, \, \Psi + \Psi \left|\Psi\right|^2\\    +
\sigma \Psi \Delta_R \left|\Psi\right|^2 - \frac{\Psi
\left|\Psi\right|^2}{R^2}=0,\label{eq:rescaling}
\end{eqnarray}
where $\sigma=7/2$. (In the limit $\sigma=0$ equation
(\ref{eq:rescaling}) reduces to equation (18) of Ref.
\cite{Kuznetsov1976}.) We will keep $\sigma$ as free parameter to
study the impact of nonlocality on the properties of 2D and 3D
Langmuir solitons and bearing in mind the other possible
applications of model equation (\ref{eq:rescaling}).

Soliton solutions of Eq. (\ref{eq:rescaling}) have a form
$\Psi(R,\tau)=\psi(R)\exp{\textrm({i}\lambda \tau)}$, where
$\lambda$ is the nonlinear frequency shift of the soliton. The
radial soliton profiles $\psi(R)$ are found from the equation
\begin{eqnarray}\nonumber
-\lambda \psi + \frac{\partial^2 \psi}{\partial R^2} +
\frac{d-1}{R} \frac{\partial \psi}{\partial R } - \frac{d-1}{R^2}
\, \, \psi + \psi \left|\psi\right|^2 \\ + \sigma \psi \Delta_R
\left|\psi\right|^2 - \frac{\psi \left|\psi\right|^2}{R^2}=0 .
\label{GNSE2D3D_Stationary}
\end{eqnarray}

We will start our consideration with 2D Langmuir solitons.
 Note the formal analogy between Eq. (\ref{GNSE2D3D_Stationary})
for radial electric field component in 2D case and NLSE for vortex
solitons $\psi(R)\exp{(\textrm{i}m\varphi)}$ with topological
charge $m=1$ (see, e.g. \cite{OurPRE03}). However, in spite of
zero value of wave intensity at the soliton center, it has no
phase dislocation and corresponds to the ground state with minimum
energy and zero angular momentum. In the considered case of
Langmuir wave structures, the phase does not depend on the radial
coordinate, thus in Eq. (\ref{GNSE2D3D_Stationary}) $\psi(R)$ may
be considered as real function.

Stationary states of Eq. (\ref{GNSE2D3D}) were investigated
analytically and numerically. Analytical approach employs
approximate variational method (see, e.g. \cite{Anderson1983})
with the normalized trial function of the form
%%%
\begin{equation}
E(r)=\sqrt{ N\mu^2/\pi}\,\,\xi\exp{(-\xi^2/2)}, \,\,\, \xi=\mu r,
 \label{TrialFunction}
\end{equation}
where the variational parameter $\mu$ characterizes the inverse
soliton's width: $\mu=1/(\sqrt{2}\,r_{\textrm{eff}})$. We have
returned to the nonscaled variables for variational analysis. The
trial function (\ref{TrialFunction}) has a correct asymptotic near
the soliton center. The Gaussian profile gives a good
approximation to soliton solutions, since, as it was argued in
\cite{BangKrolikowskiWyllerRasmussen2002} it represents an exact
solution in the limit case of strong nonlocality. The variational
parameter $\mu_0$ corresponding to stationary solutions of the
form (\ref{TrialFunction}), is readily found after standard
procedure \cite{Anderson1983}:
\begin{equation}
\mu_0^2=\mu_*^2\,(N-N_0)/N,
 \label{eq:Mu_0Sqrd}
\end{equation}
where $\mu_{*}^2=B/(4C+2\Gamma)$, $N_0=16\pi D/B$. Thus, 2D
Langmuir solitons are formed only when some threshold value of
plasmon number is exceeded (i.e. if $N>N_0$). One can see that
$\mu_0 \to \mu_*$ at $N \gg N_0$, so that the effective width is
bounded from below: $r_{\textrm{eff}} \ge \sqrt{(2C + \Gamma)/B}$,
which agrees with previous general estimate given by Eq.
(\ref{Reff_Estimat}).

The boundary-value problem described by the Eq.
(\ref{GNSE2D3D_Stationary}) with zero boundary conditions at the
soliton's center ($R = 0$) and at infinity ($R\to\infty$) was
solved numerically by the shooting method. Solitons form
two-parameter family with parameters $\sigma$ and $\lambda$. For
each given nonlocality parameter $\sigma$ we present the
dependence $N(\lambda)$ known as ``energy dispersion diagram"
(EDD), which are given in Fig. \ref{Fig2d} (a). Note that all
solitons are stable, which is similar to Vakhitov-Kolokolov
criterion \cite{VakhitovKolokolov}, since $\partial N /\partial
\lambda >0$. To excite stable two-dimensional Langmuir soliton,
the threshold value $N_0$ of input power (critical plasmon number)
should be exceeded. Variational approach predicts normalized
critical plasmon number to be $N_0=16\pi$. This is in a very good
agreement with our numerical results, where $N_0 \approx 48.3$.
Moreover, this simple variational analysis gives a good
description for all EDDs, and variational dependence $N(\lambda)$
fits better the computed one when the parameter $\sigma$ grows. It
is illustrated in the Fig. \ref{Fig2d} (a) for different $\sigma$.
One can see that at $\sigma=0$, the shape of soliton differs
sufficiently from the simple profile of the form
(\ref{TrialFunction}) [compare Fig. \ref{Fig2d} (c) with Fig.
\ref{Fig2d} (d),(e)]. The typical soliton profiles are plotted in
Fig. \ref{Fig2d} (c)-(e). At the same plasmon number, the soliton
width is larger for solutions of Eq. (\ref{GNSE2D3D_Stationary})
with nonlocal nonlinearity ($\sigma>0$) than for solutions with
$\sigma=0$.

As it was stressed above, the effective width of Langmuir wave
packet in our model is bounded from below. Figure \ref{Fig2d} (b)
represents the effective soliton width as function of plasmon
number. The $R_{\textrm{eff}}$ decays monotonically and saturates
at some nonzero minimum value $R_{\textrm{min}}$. This minimum
width increases when parameter $\sigma$ increases, as it was
estimated above [see Eq. (\ref{Reff_Estimat})]. In the dimensional
variables, the minimum diameters of 2D Langmuir structures are of
order $10 r_D$ which is in a good agreement with those observed
experimentally in \cite{Eggleston1982}.

We have performed similar analytical and numerical study for 3D
radially-symmetric solitons. The corresponding $N(\lambda)$ and
$R_{\textrm{eff}}(N)$ dependencies are shown in Fig. \ref{Fig3d}
(a), (b). In the 3D case, plasmon number depends on nonlinear
frequency shift $\lambda$ non-monotonically, and the only stable
soliton branch corresponds to $\partial N/\partial \lambda>0$, as
it follows from Vakhitov-Kolokolov criterion. Therefore, in the 3D
case, the threshold plasmon number $N_{cr}$ needed to excite
stable Langmuir soliton should be obtained from condition
$\partial N/\partial \lambda =0$. Similarly to the 2D case, the
effective width $R_{\textrm{eff}}$ of stable solitons decays
monotonically when plasmon number $N$ grows, and $R_{\textrm{eff}}
\to R_{\textrm{min}}$ at $N \gg N_{cr}$. We have found the
threshold value to be of order $N_{cr} \approx 850$.
%---------------------------------------------

We have investigated evolution of radially symmetric Langmuir
structures numerically in the framework of Eq.
(\ref{eq:rescaling}). The integration time step was splitted into
the linear and nonlinear parts, and both of them were performed in
the radial coordinate domain. During the simulations, the
conservation of integrals (\ref{PlNum2D}) and (\ref{Hamilt2D}) has
been verified, and if the change of any integral exceeded 1\%, the
modelling was stopped. We used two types of initial conditions:
(i) perturbed stationary soliton solution of Eq.
(\ref{GNSE2D3D_Stationary}) which was found numerically, and (ii)
the Gaussian-like profile. At the boundaries we assumed that
$\partial\left[\Psi(R,t)/R\right]/\partial R=0$ at $R\to 0$ and
$\Psi(R,t)=0$ at $R\to \infty$.

\begin{figure*}[hbt]
\begin{centering}
\includegraphics[width=\textwidth]{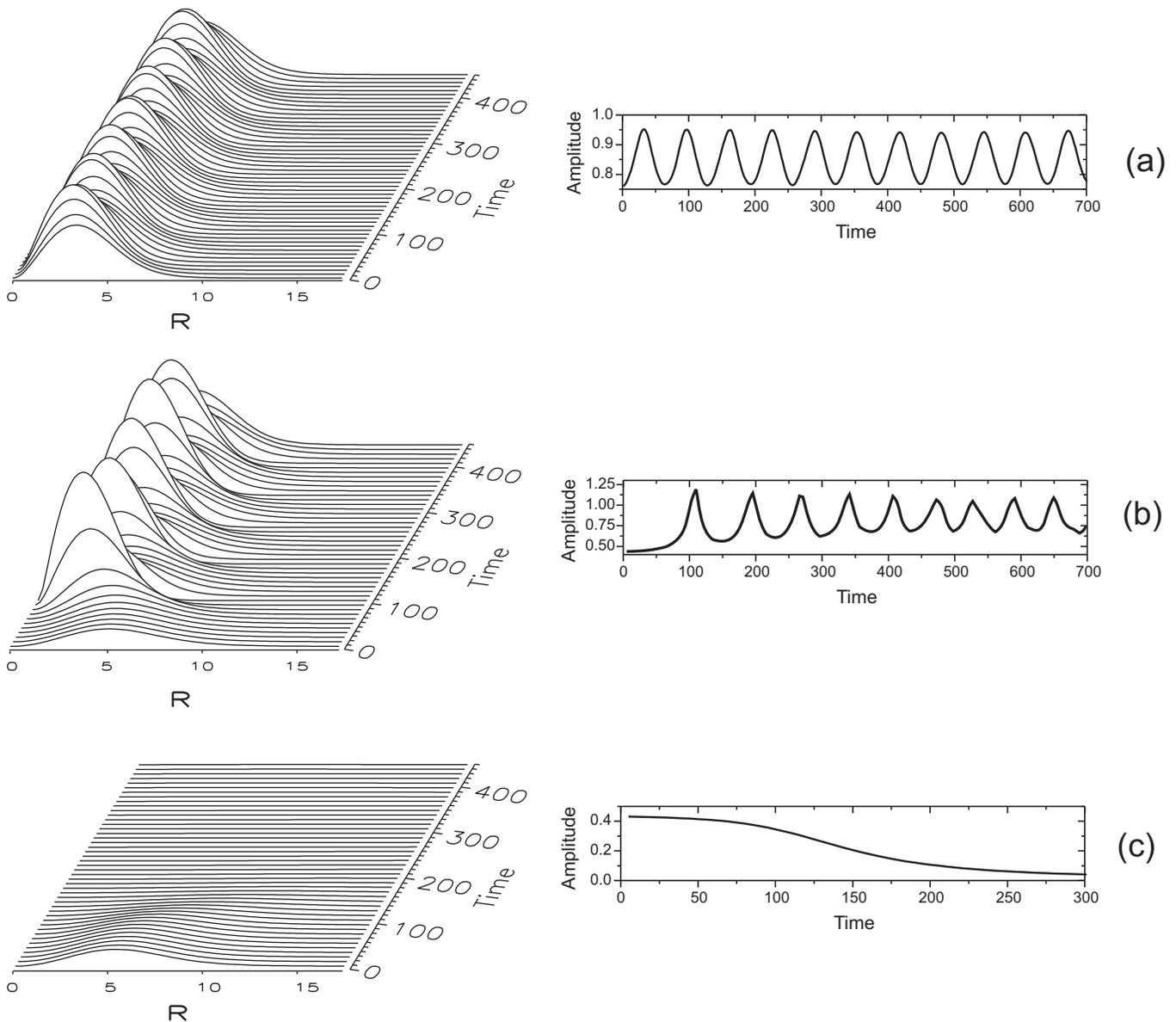}
 \caption{Typical examples of evolution of
 3D Langmuir wave packets ($\sigma=1.5$).
 Left panels
 represent distributions of wave intensity $|\Psi(R,t)|^2$. Right panels
 show wave packet's amplitude $\Psi_{\textrm{max}}$ vs time. (a)
 perturbed stable soliton with $\lambda=0.1$; perturbed
 soliton from unstable branch at $\lambda=0.03$ and
 different perturbations: (b) demonstrates quasiperiodical dynamics;
 (c) spreads out.}
  \label{Fig:Dynamics}
\end{centering}
\end{figure*}

Stability properties of Langmuir solitons were found to be in a
very good agreement with our analytical predictions. Any localized
wave packet having the number of quanta below the threshold value
$N<N_{cr}$ always spreads out, while an initially more intense
packet with $N>N_{cr}$ may form a localized structure after
irradiating a portion of plasmons. The dynamics of intense wave
packet with $N>N_{cr}$ is very intriguing. Figure
\ref{Fig:Dynamics} represents the different evolution scenarios of
perturbed 3D Langmuir solitons. The amplitude and effective width
of slightly perturbed soliton solution belonging to the stable
branch (with $\partial N/\partial \lambda>0$) oscillate in time as
it is seen in Fig. \ref{Fig:Dynamics} (a). However, when
perturbed, an unstable stationary solution (having $\partial
N/\partial \lambda<0$) manifests different evolution pattern: it
can either monotonically spread out [see Fig. \ref{Fig:Dynamics}
(c)] or develop quasiperiodical motion [see Fig.
\ref{Fig:Dynamics} (b)], depending on the initial perturbation.
The latter quasiperiodical behavior resembles the oscillations
between two opposite extreme states discussed in
\cite{BangKrolikowskiWyllerRasmussen2002,
PerezGarciaKonotopPRE2000}. The initial increase of wave packet's
intensity is followed by quasiperiodical oscillations with rather
large amplitude. Thus, the additional nonlinearities actually
prevent catastrophic collapse of any 3D Langmuir wave packets. The
similar regime is known  as "frustrated collapse"
\cite{PerezGarciaKonotopPRE2000} in theory of Bose-Einstein
condensates with nonlocal nonlinear interaction.
%---------------------------------------------
%---------------------------------------------

Our considerations are in a qualitative agreement with
experimental observations of Langmuir wave structures in
unmagnetized laboratory plasmas \cite{Wong1984, Wong1985}. On the
slow (subsonic) stage of evolution, which was observed at time
$t>50\omega_{Pi}^{-1}$, the field envelope almost stopped its
radial contraction and the size remained approximately unchanged
($r$ is of order of few tens of $r_D$). It is important to note
that according to \cite{Wong1984,Wong1985} at this stage of
Langmuir structure evolution, the beam-wave resonance is detuned
and the beam decouples from the wave. Nevertheless, the common
Zakharov set of equations describes field evolution only for
rather short time ($t<50\omega_{Pi}^{-1}$) and then its prediction
drastically deviates from experimental facts because the wave
packet becomes too intense (at peak $E_{max}^2/4\pi nT\sim 1$) and
too narrow. As it was demonstrated above, the electron-electron
nonlinearity plays a crucial role in Langmuir structure behavior
and it gives the qualitative explanation of saturation of the wave
packet contraction. The dissipative effects such as Landau and
transit-time damping seems to be negligible for structures of
several tens Debye's radius characteristic size. Certainly, the
wave packet may interact with high-energy electrons, therefore,
after a long time (of order of several hundreds of
$\omega_{Pi}^{-1}$), it may eventually damp due to wave
absorption. Our considerations are valid at stage when nonlocal
nonlinearities come into play and considerably slow down the wave
packet contraction but the wave absorption is still not essential.

In conclusion, we have performed analytical and numerical studies
of spatial (2D and 3D) Langmuir solitons in the framework of model
based on generalized nonlinear Sch\"odinger equation including
both local and nonlocal electron-electron nonlinearities. Their
influence on intense and narrow Langmuir wave packets are of the
same order, and both nonlinearities should be taken into account
simultaneously. Any of them is able to arrest the Langmuir
collapse. Both nonlinearities lead to the saturation of soliton
width with an increase of the energy, but quantitatively the
effect of nonlocal nonlinearity is more significant. All 2D
Langmuir solitary structures are stable, while in 3D case, two
soliton branches coexist, one is stable and the other is unstable.
When perturbed, stable solitons demonstrate centrosymmetric
oscillations. As for 3D solitons from the unstable branch, they
may either spread out or oscillate quasiperiodically depending on
perturbation applied.

%In 3D, this simple model does not predict
%jphysically realistic spherically symmetric Langmuir solitons.

\end{document}